\documentclass[12pt,a4paper]{article}

\usepackage{ifthen} 
\usepackage{subfigure}
\newboolean{pdflatex}
\setboolean{pdflatex}{true} 

\newboolean{articletitles}
\setboolean{articletitles}{true} 

\newboolean{uprightparticles}
\setboolean{uprightparticles}{false} 

\usepackage{microtype}
\usepackage{lineno}  
\usepackage{xspace} 

\usepackage{graphicx}  
\usepackage{color}
\usepackage{colortbl}
\graphicspath{{./figs/}} 

\usepackage{amsmath} 
\usepackage{amssymb}
\usepackage{amsfonts}
\usepackage{upgreek} 

\newcommand*\patchAmsMathEnvironmentForLineno[1]{%
\expandafter\let\csname old#1\expandafter\endcsname\csname #1\endcsname
\expandafter\let\csname oldend#1\expandafter\endcsname\csname
end#1\endcsname
 \renewenvironment{#1}%
   {\linenomath\csname old#1\endcsname}%
   {\csname oldend#1\endcsname\endlinenomath}%
}
\newcommand*\patchBothAmsMathEnvironmentsForLineno[1]{%
  \patchAmsMathEnvironmentForLineno{#1}%
  \patchAmsMathEnvironmentForLineno{#1*}%
}
\AtBeginDocument{%
\patchBothAmsMathEnvironmentsForLineno{equation}%
\patchBothAmsMathEnvironmentsForLineno{align}%
\patchBothAmsMathEnvironmentsForLineno{flalign}%
\patchBothAmsMathEnvironmentsForLineno{alignat}%
\patchBothAmsMathEnvironmentsForLineno{gather}%
\patchBothAmsMathEnvironmentsForLineno{multline}%
}

\usepackage{hyperref}    
\usepackage[all]{hypcap} 

\def\lhcb {\mbox{LHCb}\xspace}
\def\ux85 {\mbox{UX85}\xspace}

\ifthenelse{\boolean{uprightparticles}}%
{

 \def\Pmu         {\ensuremath{\upmu}\xspace}

 \def\Pchi        {\ensuremath{\upchi}\xspace}                 
 \def\Ppsi        {\ensuremath{\uppsi}\xspace}

 \def\PDelta      {\ensuremath{\Delta}\xspace}                 
 \def\PXi      {\ensuremath{\Xi}\xspace}                 
 \def\PLambda      {\ensuremath{\Lambda}\xspace}                 
 \def\PSigma      {\ensuremath{\Sigma}\xspace}                 
 \def\POmega      {\ensuremath{\Omega}\xspace}                 
 \def\PUpsilon      {\ensuremath{\Upsilon}\xspace}                 
 
 \def\PB      {\ensuremath{\mathrm{B}}\xspace}                 
                  
 \def\PD      {\ensuremath{\mathrm{D}}\xspace}

 \def\PJ      {\ensuremath{\mathrm{J}}\xspace}                 
 \def\PK      {\ensuremath{\mathrm{K}}\xspace}

 \def\Pb      {\ensuremath{\mathrm{b}}\xspace}                 
 \def\Pc      {\ensuremath{\mathrm{c}}\xspace}

 \def\Pi      {\ensuremath{\mathrm{i}}\xspace}

}
{

 \def\Pmu         {\ensuremath{\mu}\xspace}

 \def\Pchi        {\ensuremath{\chi}\xspace}                 
 \def\Ppsi        {\ensuremath{\psi}\xspace}                 
                  
 \mathchardef\PDelta="7101
 \mathchardef\PXi="7104
 \mathchardef\PLambda="7103
 \mathchardef\PSigma="7106
 \mathchardef\POmega="710A
 \mathchardef\PUpsilon="7107
                  
 \def\PB      {\ensuremath{B}\xspace}                 
                  
 \def\PD      {\ensuremath{D}\xspace}

 \def\PJ      {\ensuremath{J}\xspace}                 
 \def\PK      {\ensuremath{K}\xspace}

 \def\Pb      {\ensuremath{b}\xspace}                 
 \def\Pc      {\ensuremath{c}\xspace}

 \def\Pi      {\ensuremath{i}\xspace}

}

\def\mumu       {\ensuremath{\Pmu^+\Pmu^-}\xspace}

\def\cquark    {\ensuremath{\Pc}\xspace}

\def\bquark    {\ensuremath{\Pb}\xspace}

\def\kaon  {\ensuremath{\PK}\xspace}
  \def\Kbar  {\kern 0.2em\overline{\kern -0.2em \PK}{}\xspace}

\def\Kz    {\ensuremath{\kaon^0}\xspace}
\def\Kzb   {\ensuremath{\Kbar^0}\xspace}
\def\KzKzb {\ensuremath{\Kz \kern -0.16em \Kzb}\xspace}
\def\Kp    {\ensuremath{\kaon^+}\xspace}
\def\Km    {\ensuremath{\kaon^-}\xspace}

\def\KpKm  {\ensuremath{\Kp \kern -0.16em \Km}\xspace}

  \def\Dbar    {\kern 0.2em\overline{\kern -0.2em \PD}{}\xspace}
\def\D       {\ensuremath{\PD}\xspace}

\def\Dz      {\ensuremath{\D^0}\xspace}
\def\Dzb     {\ensuremath{\Dbar^0}\xspace}
\def\DzDzb   {\ensuremath{\Dz {\kern -0.16em \Dzb}}\xspace}
\def\Dp      {\ensuremath{\D^+}\xspace}
\def\Dm      {\ensuremath{\D^-}\xspace}

\def\DpDm    {\ensuremath{\Dp {\kern -0.16em \Dm}}\xspace}

  \def\Bbar    {\kern 0.18em\overline{\kern -0.18em \PB}{}\xspace}

\def\jpsi     {\ensuremath{{\PJ\mskip -3mu/\mskip -2mu\Ppsi\mskip 2mu}}\xspace}
\def\psitwos  {\ensuremath{\Ppsi{(2S)}}\xspace}

  \def\Y#1S{\ensuremath{\PUpsilon{(#1S)}}\xspace}

\def\chic  {\ensuremath{\Pchi_{c}}\xspace}

\def\Lbar {\ensuremath{\kern 0.1em\overline{\kern -0.1em\PLambda}}\xspace}

\def\AT#1     {\ensuremath{A_{\mathrm{T}}^{#1}}\xspace}           

\def\C#1      {\ensuremath{\mathcal{C}_{#1}}\xspace}                       
\def\Cp#1     {\ensuremath{\mathcal{C}_{#1}^{'}}\xspace}                    
\def\Ceff#1   {\ensuremath{\mathcal{C}_{#1}^{\mathrm{(eff)}}}\xspace}        
\def\Cpeff#1  {\ensuremath{\mathcal{C}_{#1}^{'\mathrm{(eff)}}}\xspace}       
\def\Ope#1    {\ensuremath{\mathcal{O}_{#1}}\xspace}                       
\def\Opep#1   {\ensuremath{\mathcal{O}_{#1}^{'}}\xspace}                    

\newcommand{\tev}{\ensuremath{\mathrm{\,Te\kern -0.1em V}}\xspace}
\newcommand{\gev}{\ensuremath{\mathrm{\,Ge\kern -0.1em V}}\xspace}
\newcommand{\mev}{\ensuremath{\mathrm{\,Me\kern -0.1em V}}\xspace}
\newcommand{\kev}{\ensuremath{\mathrm{\,ke\kern -0.1em V}}\xspace}
\newcommand{\ev}{\ensuremath{\mathrm{\,e\kern -0.1em V}}\xspace}
\newcommand{\gevc}{\ensuremath{{\mathrm{\,Ge\kern -0.1em V\!/}c}}\xspace}
\newcommand{\mevc}{\ensuremath{{\mathrm{\,Me\kern -0.1em V\!/}c}}\xspace}
\newcommand{\gevcc}{\ensuremath{{\mathrm{\,Ge\kern -0.1em V\!/}c^2}}\xspace}
\newcommand{\gevgevcccc}{\ensuremath{{\mathrm{\,Ge\kern -0.1em V^2\!/}c^4}}\xspace}
\newcommand{\mevcc}{\ensuremath{{\mathrm{\,Me\kern -0.1em V\!/}c^2}}\xspace}

\def\mum  {\ensuremath{\,\upmu\rm m}\xspace}

\def\invpb {\ensuremath{\mbox{\,pb}^{-1}}\xspace}

\def\gsim{{~\raise.15em\hbox{$>$}\kern-.85em
          \lower.35em\hbox{$\sim$}~}\xspace}
\def\lsim{{~\raise.15em\hbox{$<$}\kern-.85em
          \lower.35em\hbox{$\sim$}~}\xspace}

\def\sqs   {\ensuremath{\protect\sqrt{s}}\xspace}

\def\pt         {\mbox{$p_{\rm T}$}\xspace}

\def\tell1  {TELL1\xspace}
\def\ukl1   {UKL1\xspace}

\usepackage{cite} 
\usepackage{mciteplus}

\begin{document}

\renewcommand{\thefootnote}{\fnsymbol{footnote}}
\setcounter{footnote}{1}


\begin{titlepage}
\pagenumbering{roman}

\vspace*{-1.5cm}
\centerline{\large EUROPEAN ORGANIZATION FOR NUCLEAR RESEARCH (CERN)}
\vspace*{1.5cm}
\hspace*{-0.5cm}
\begin{tabular*}{\linewidth}{lc@{\extracolsep{\fill}}r}
\ifthenelse{\boolean{pdflatex}}
{\vspace*{-2.7cm}\mbox{\!\!\!\includegraphics[width=.14\textwidth]{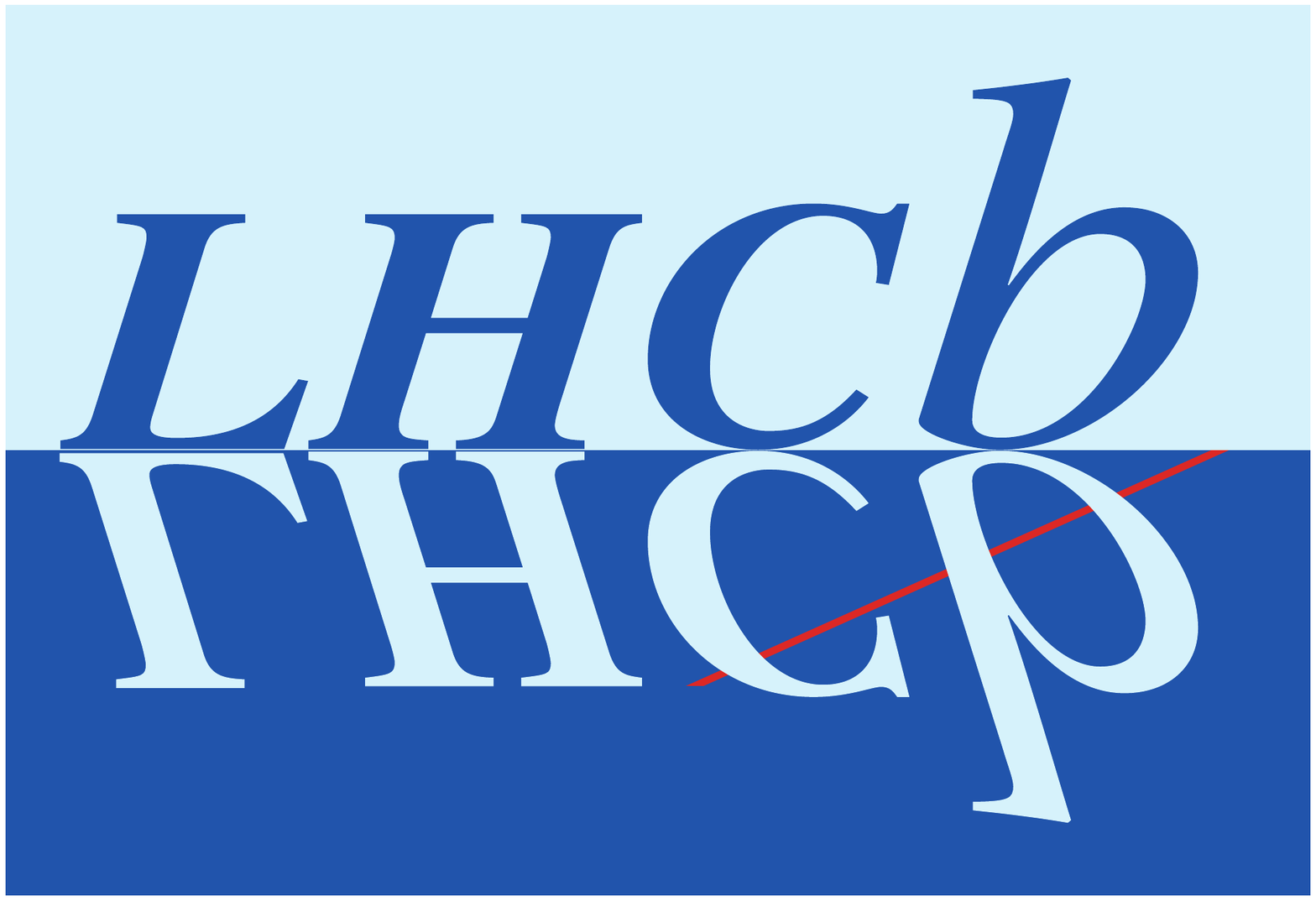}} & &}%
{\vspace*{-1.2cm}\mbox{\!\!\!\includegraphics[width=.12\textwidth]{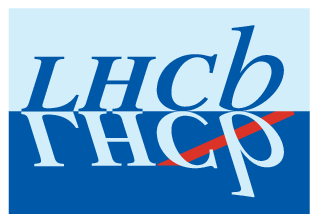}} & &}%
\\
 & & CERN-PH-EP-2013-005 \\  
 & & LHCb-PAPER-2012-044 \\  
 & & 29 January 2013 \\ 
 & & \\
\end{tabular*}

\vspace*{2.5cm}

{\bf\boldmath\huge
\begin{center}
  Exclusive $\jpsi$ and $\psitwos$ production in \textit{pp} collisions at $\sqs = 7$ \tev
\end{center}
}

\vspace*{1.0cm}

\begin{center}
The LHCb collaboration\footnote{Authors are listed on the following pages.}
\end{center}

\vspace*{1.0cm}

\begin{abstract}
  \noindent

Exclusive $\jpsi$ and $\psitwos$ vector meson production has been observed in the dimuon channel using the $\lhcb$ detector. 
The cross-section times branching fractions 
to two muons with pseudorapidities between 2.0 and 4.5 are measured
to be
\begin{equation*}
\sigma_{pp\rightarrow \jpsi (\rightarrow \mu^{+} \mu^{-})} (2.0 <\eta_{\mu^{\pm}}< 4.5) = 307 \pm 21 \pm 36~\text{pb}, 
\end{equation*}
\begin{equation*}
\sigma_{pp\rightarrow \psitwos (\rightarrow \mu^{+} \mu^{-})} (2.0 <\eta_{\mu^{\pm}}< 4.5) = 7.8 \pm 1.3 \pm 1.0~\text{pb}, 
\end{equation*}
  \noindent
where the first uncertainties are statistical and the second are systematic.
The measurements are found to be in good agreement with results from previous experiments and theoretical predictions. 
The $\jpsi$ photoproduction cross-section has been  measured as a function of the photon-proton centre-of-mass energy. The results are consistent with measurements obtained at HERA and confirm a similar power law behaviour for the photoproduction cross-section.
 
\end{abstract}

\vspace*{1.0cm}

\begin{center}
  To be submitted to Journal of Physics G
  \end{center}

\vspace{\fill}

{\footnotesize
\centerline{\copyright~CERN on behalf of the \lhcb collaboration, license \href{http://creativecommons.org/licenses/by/3.0/}{CC-BY-3.0}.}}
\vspace*{2mm}

\end{titlepage}


\newpage
\setcounter{page}{2}
\mbox{~}
\newpage

\centerline{\large\bf LHCb collaboration}
\begin{flushleft}
\small
R.~Aaij$^{38}$, 
C.~Abellan~Beteta$^{33,n}$, 
A.~Adametz$^{11}$, 
B.~Adeva$^{34}$, 
M.~Adinolfi$^{43}$, 
C.~Adrover$^{6}$, 
A.~Affolder$^{49}$, 
Z.~Ajaltouni$^{5}$, 
J.~Albrecht$^{9}$, 
F.~Alessio$^{35}$
% ...

{\footnotesize \it
% ...
$ ^{5}$Clermont Universit\'{e}, Universit\'{e} Blaise Pascal, CNRS/IN2P3, LPC, Clermont-Ferrand, France\\
$ ^{6}$CPPM, Aix-Marseille Universit\'{e}, CNRS/IN2P3, Marseille, France\\
% ...
$ ^{9}$Fakult\"{a}t Physik, Technische Universit\"{a}t Dortmund, Dortmund, Germany\\
% ...
$ ^{11}$Physikalisches Institut, Ruprecht-Karls-Universit\"{a}t Heidelberg, Heidelberg, Germany\\
% ...
$ ^{33}$Universitat de Barcelona, Barcelona, Spain\\
$ ^{34}$Universidad de Santiago de Compostela, Santiago de Compostela, Spain\\
$ ^{35}$European Organization for Nuclear Research (CERN), Geneva, Switzerland\\
% ...
$ ^{38}$Nikhef National Institute for Subatomic Physics, Amsterdam, The Netherlands\\
% ...
$ ^{43}$H.H. Wills Physics Laboratory, University of Bristol, Bristol, United Kingdom\\
% ...
$ ^{49}$Oliver Lodge Laboratory, University of Liverpool, Liverpool, United Kingdom\\
% ...
$ ^{n}$LIFAELS, La Salle, Universitat Ramon Llull, Barcelona, Spain\\
% ...
}
\end{flushleft}

\cleardoublepage

\renewcommand{\thefootnote}{\arabic{footnote}}
\setcounter{footnote}{0}

\pagestyle{plain} 
\setcounter{page}{1}
\pagenumbering{arabic}
\interfootnotelinepenalty=10000


\section{Introduction}

Exclusive vector meson production through photoproduction, $\gamma p\rightarrow V p$, has attracted much interest both experimentally and theoretically as it provides a rich testing ground for QCD.
At sufficiently high meson masses, perturbative QCD (pQCD) can be used to predict the production cross-section~\cite{pQCDrise}, but as the
masses decrease this approach ceases to work due to non-perturbative effects.  
Light vector meson production is best described by Regge theory~\cite{Regge}, which models the process using soft pomeron exchange. This theory predicts a cross-section which is almost flat with respect to  $W$, the photon-proton centre-of-mass energy.  
In contrast, the cross-section for exclusive $\jpsi$ production has been observed to have a strong power law dependence on $W$~\cite{cH1b, zeus}. This feature is described in pQCD by a hard pomeron (or two gluon exchange) and the fact that the gluon density in the proton increases rapidly with decreasing $x$, the fractional momentum of the proton carried by the parton~\cite{pQCDrise}.
A consistent description of the transition regime between perturbative and non-perturbative QCD is a challenge.

The cross-section for exclusive $\jpsi$ and $\psitwos$ production is calculable within pQCD though with large uncertainties. The simplest leading-order calculation is proportional to $[\alpha_S(m_{V}^2/4)xg(x,m_{V}^2/4)]^2$, where $m_{V}$ is the vector meson mass, $\alpha_S(m_{V}^2/4)$ is the strong coupling constant and $g(x,m_{V}^2/4)$ is the gluon parton density function (PDF) evaluated at the relevant scale, $m_{V}^2/4$. Hence a measurement of these processes at the LHC can constrain the gluon PDF~\cite{pQCDrise}.  

More exotic effects can also be searched for in this process. The forward acceptance of the LHCb detector is sensitive to $W$ values between 10 and 2000 GeV and allows the
gluon density to be probed down to $x = 5\times10^{-6}$~\cite{PDFatLHC}, lower than any previous experiment at a scale of a few $\gev$.
At a sufficiently low value of $x$, the strong growth of the gluon density with decreasing $x$ is expected to be restrained by gluon recombination~\cite{Saturation}. This saturation effect could be visible as a reduction in the photoproduction cross-section with respect to the power law behaviour at $W$ values accessible at the LHC~\cite{cMW}. 
Exclusive vector meson production also provides a promising channel to investigate the existence of the odderon~\cite{Odd}.

The HERA experiments measured $\jpsi$ photoproduction using electron-proton collisions in the range 20 $\le W \le$ 305 GeV~\cite{cH1b, zeus}.
The data are consistent with a power law dependence of the form $\sigma= a(W/1\gev)^\delta$ with $a=3$ nb and $\delta=0.72$~\cite{cMW}.
The CDF experiment measured exclusive $\jpsi$ production in proton-antiproton collisions in a similar kinematic range as HERA and found consistent results~\cite{cCDF}.  

This paper presents measurements of the
cross-section times branching fractions
for exclusive $\jpsi$ and $\psitwos$ mesons to produce
two muons in the pseudorapidity range $2.0<\eta_{\mu^{\pm}}<4.5$ at a proton-proton centre-of-mass energy of $\sqrt{s} =7$ $\tev$. 
The results are compared with previous experiments and a number of theoretical models. 
The \textsc{SuperChic}~\cite{cSUPERC} and \textsc{Starlight}~\cite{klein04} models 
use a parameterisation of the HERA results to predict the photoproduction cross-section at the LHC.  
Motyka and Watt~\cite{cMW} use an equivalent photon approximation combined with a dipole model, which is found to reproduce the main features of the HERA data.
Gon\c{c}alves and Machado~\cite{Machado} use a colour dipole approach and the Colour Glass
Condensate formalism which also agrees with the HERA data.
Sch\"afer and Szczurek~\cite{schaefer} use an explicit dynamical model of the photoproduction amplitude, which is evaluated in terms of an unintegrated gluon
distribution.  This model predicts a cross-section which is higher than that observed at HERA.

\section{Detector and data samples}

The \lhcb detector~\cite{Alves:2008zz} is a single-arm forward
spectrometer covering the \mbox{pseudorapidity} range $2<\eta <5$ (forward region), designed
for the study of particles containing \bquark or \cquark quarks. 
The detector includes a high precision tracking system consisting of a
silicon-strip vertex detector (VELO) surrounding the proton-proton interaction region,
a large-area silicon-strip detector located upstream of a dipole
magnet with a bending power of about $4{\rm\,Tm}$, and three stations
of silicon-strip detectors and straw drift-tubes placed
downstream. 
The combined tracking system has a momentum resolution
$\Delta p/p$ that varies from 0.4$\%$ at 5$\gevc$ to 0.6$\%$ at 100$\gevc$,
and an impact parameter resolution of 20\mum for tracks with high
transverse momentum. Charged hadrons are identified using two
ring-imaging Cherenkov detectors. Photon, electron and hadron
candidates are identified by a calorimeter system consisting of
scintillating-pad (SPD) and pre-shower detectors, an electromagnetic
calorimeter and a hadronic calorimeter. The SPD also provides a measurement of the charged particle multiplicity in an event.
Muons are identified by a muon system composed of alternating layers of iron and multiwire
proportional chambers. The trigger consists of a hardware stage, based
on information from the calorimeter and muon systems, followed by a
software stage which applies a full event reconstruction.
The VELO also has sensitivity to charged particles with momenta above $\sim$100$\mevc$ in the \mbox{pseudorapidity} range
 $-3.5<\eta <-1.5$ (backward region) and extends the sensitivity of the forward region to $1.5<\eta<5$.

The $\jpsi$ and $\psitwos$ mesons are identified through their decay to two muons. 
The protons are only marginally deflected and remain undetected inside the beam pipe. 
Therefore the signature for exclusive vector meson production is an event containing two muons and no other activity. 
The analysis is performed using 36 $\invpb$ of proton-proton collision data collected at LHCb in 2010. 
In this data taking period, the average number of interactions per bunch crossing varied up to a maximum of about 2.5, with most luminosity accumulated at larger values.
 In order to identify exclusive candidates, the analysis is restricted to events with a single interaction. 

Dedicated Monte Carlo (MC) generators have been used to produce signal and background events which are passed through 
a \textsc{GEANT4}~\cite{geant} based detector simulation, the trigger emulation and the event reconstruction chain of the $\lhcb$ experiment.
Two generators have been used to produce samples of exclusive $\jpsi$ and $\psitwos$: \textsc{Starlight}~\cite{klein04}
and \textsc{SuperChic}~\cite{cSUPERC}. A sample of $\chic$ production by double pomeron fusion, which forms a background for the $\jpsi$ analysis, has been produced with \textsc{SuperChic}. 

\section{Event selection}

The hardware trigger used in this analysis requires a single muon track with transverse momentum $\pt$ $>$ 400$\mevc$, or two muon tracks 
both with $\pt$ $>$ 80$\mevc$, in coincidence with a low SPD multiplicity ($<$ 20 hits).
The software trigger used requires either a dimuon with invariant mass greater than 2.9$\gevcc$, or a dimuon with invariant mass greater 
than 1$\gevcc$ with a distance of closest approach of the two muons to each other less than 150$\mum$ and a dimuon $\pt$ less than 900$\mevc$.

\begin{figure}[t]
\begin{center}$
\begin{array}{cc}
 \includegraphics[width=0.5\linewidth]{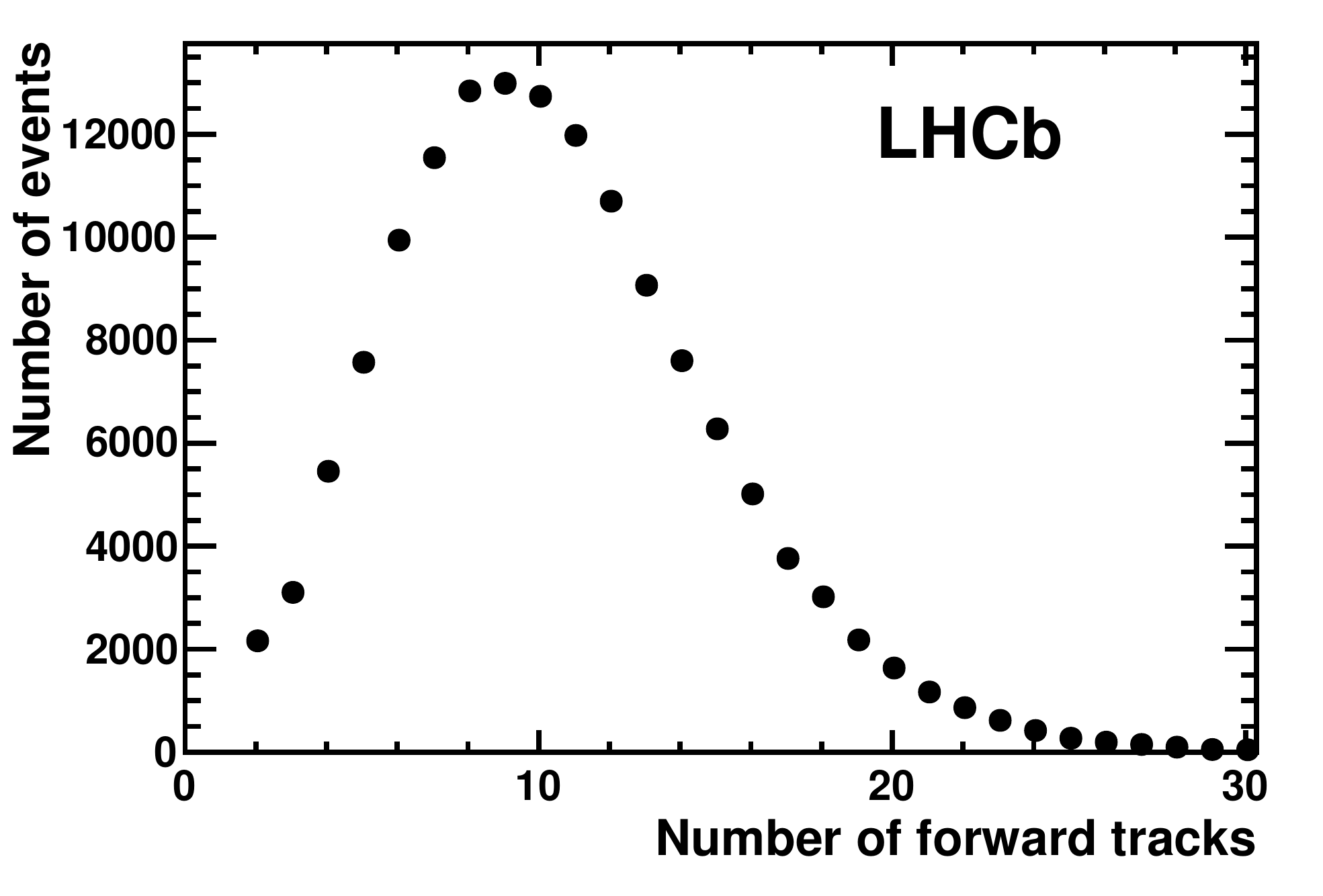} &
  \includegraphics[width=0.5\linewidth]{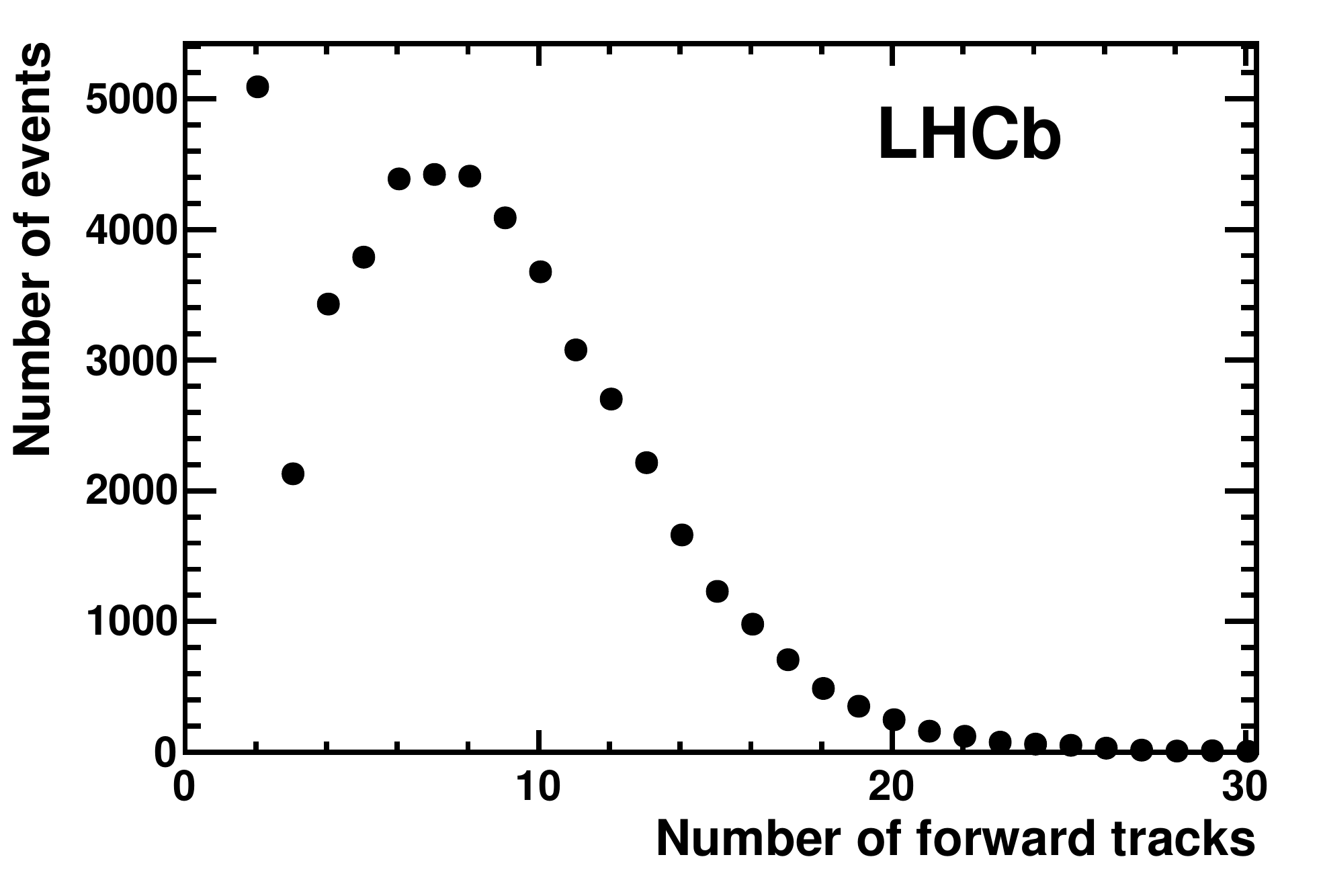} \\
     \vspace*{-1.0cm}
\end{array}$
\end{center}
\small
\caption{
Number of tracks in the forward region for dimuon triggered events which in the backward region have 
(left) one or more tracks or (right) no tracks.
}
\label{fig:TrackDist}
\end{figure}

The selection of exclusive events begins with the requirement of two reconstructed muons in the forward region. 
It is also required that there are no other tracks and no photons in the detector. Thus rapidity gaps, regions devoid of reconstructed charged and neutral objects, are defined. 
Requiring just two tracks, produced from a single particle decay,
ensures two rapidity gaps which sum to 3.5 units in the forward region.
An additional rapidity gap is obtained by requiring that there are no tracks in the backward region.
The VELO is able to exclude tracks within a certain rapidity gap depending on the $z$ position from which the tracks originate and the event topology.  The mean backward rapidity gap where tracks are excluded is 1.7 with a root mean square of about 0.5.
Figure~\ref{fig:TrackDist} shows that requiring a backward rapidity gap affects the distribution of tracks in the forward region; with a backward rapidity gap required, there is a clear peak at precisely two forward tracks. 
These are candidates for exclusive production. 

Both muons are required to be in the pseudorapidity range $2.0<\eta_{\mu^{\pm}}<4.5$. 
Muon pair candidates, having invariant masses within 65$\mevcc$ of the known $\jpsi$ and $\psitwos$ mass values~\cite{cPDG}, are selected.

\subsection{Non-resonant background determination} 

\begin{figure}[t]
\begin{center}$
\begin{array}{cc}
  \includegraphics[width=0.5\linewidth]{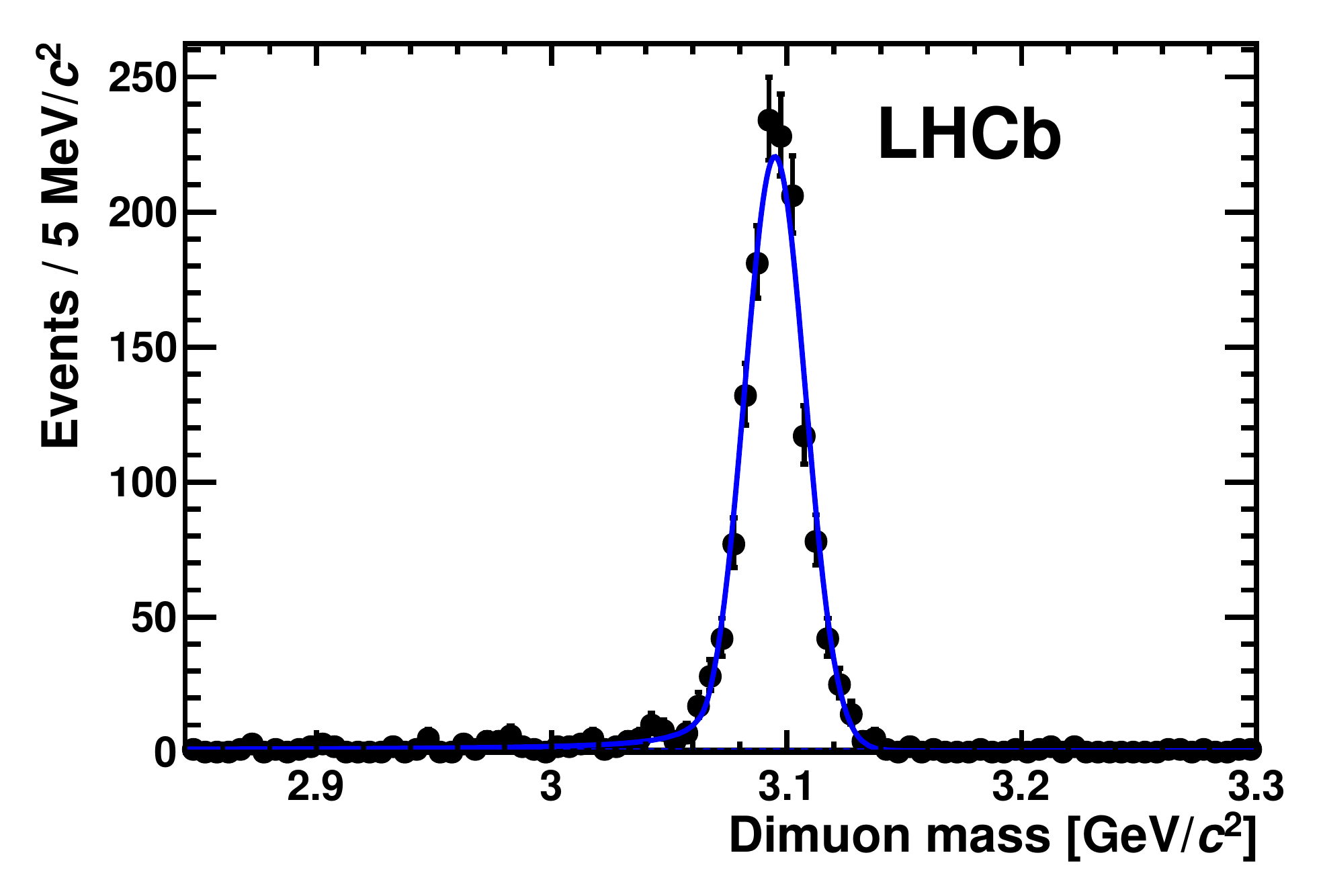} &
  \includegraphics[width=0.5\linewidth]{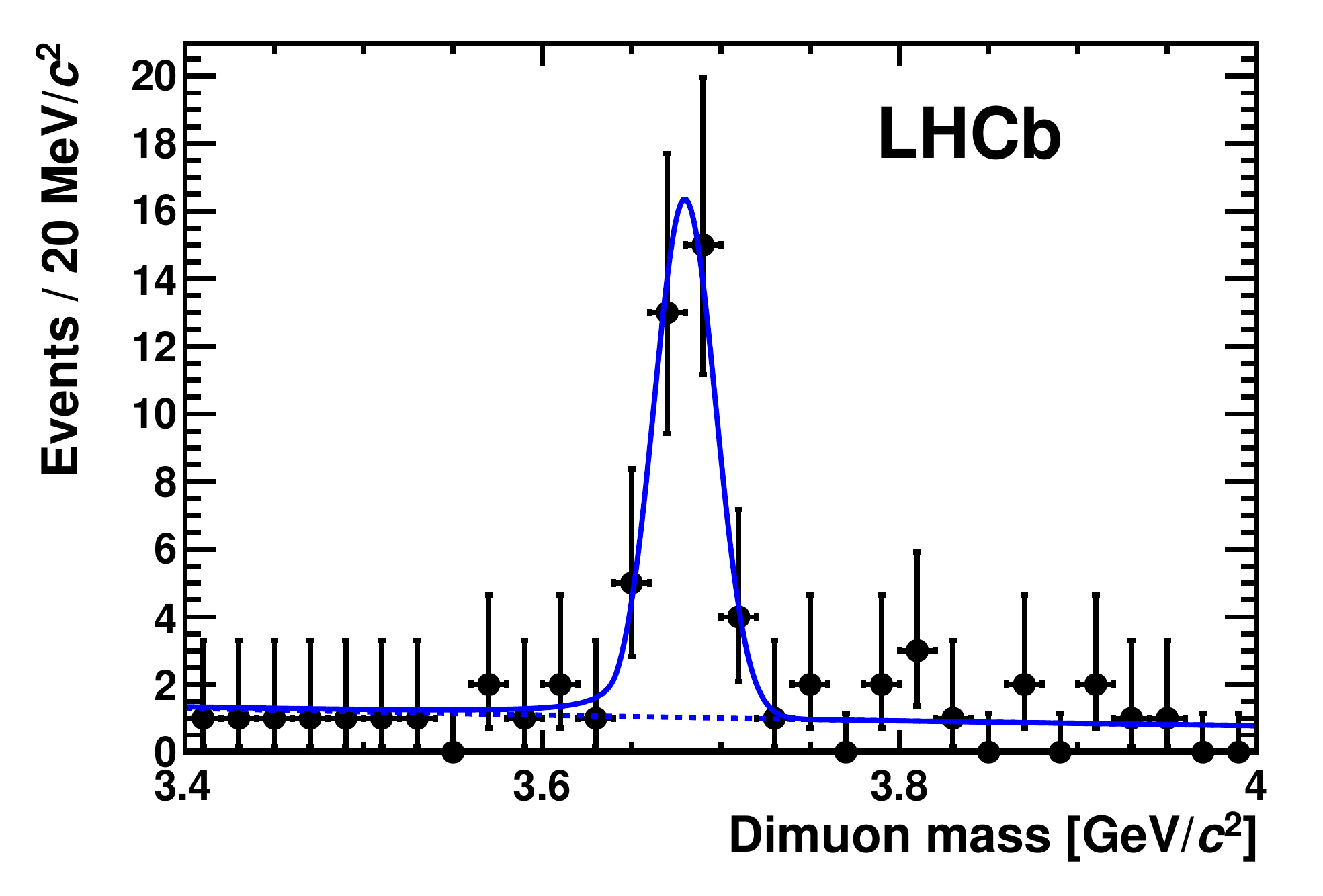} \\ 
     \vspace*{-1.0cm}
\end{array}$
\end{center}
\small
\caption{
Invariant mass distributions in the regions of (left) the $\jpsi$ and (right) $\psitwos$  mass peaks for events with exactly two tracks, no photons and a dimuon with $\pt$ below 900$\mevc$. 
The overall fits to the data are shown by the full curves while the dashed curves show the background contributions.  }
\label{fig:JPsiMass}
\end{figure}

A number of background processes have been considered including a non-resonant contribution, inclusive prompt charmonium production, inelastic photoproduction and exclusive $\chic$ and $\psitwos \rightarrow \jpsi + X$ productions. 
The non-resonant background is evaluated by fitting the dimuon invariant-mass distribution, parameterizing the resonances with a Crystal Ball function~\cite{CBall} and the continuum with an exponential function. 
Figure~\ref{fig:JPsiMass} displays the fit results. The non-resonant background is estimated to account for (0.8 $\pm$ 0.1)$\%$ and \mbox{(16 $\pm$ 3)$\%$} of the events within 65$\mevcc$ of the known $\jpsi$ and $\psitwos$ mass values, respectively.

\begin{figure}
\begin{center}
\includegraphics[width=0.9\linewidth]{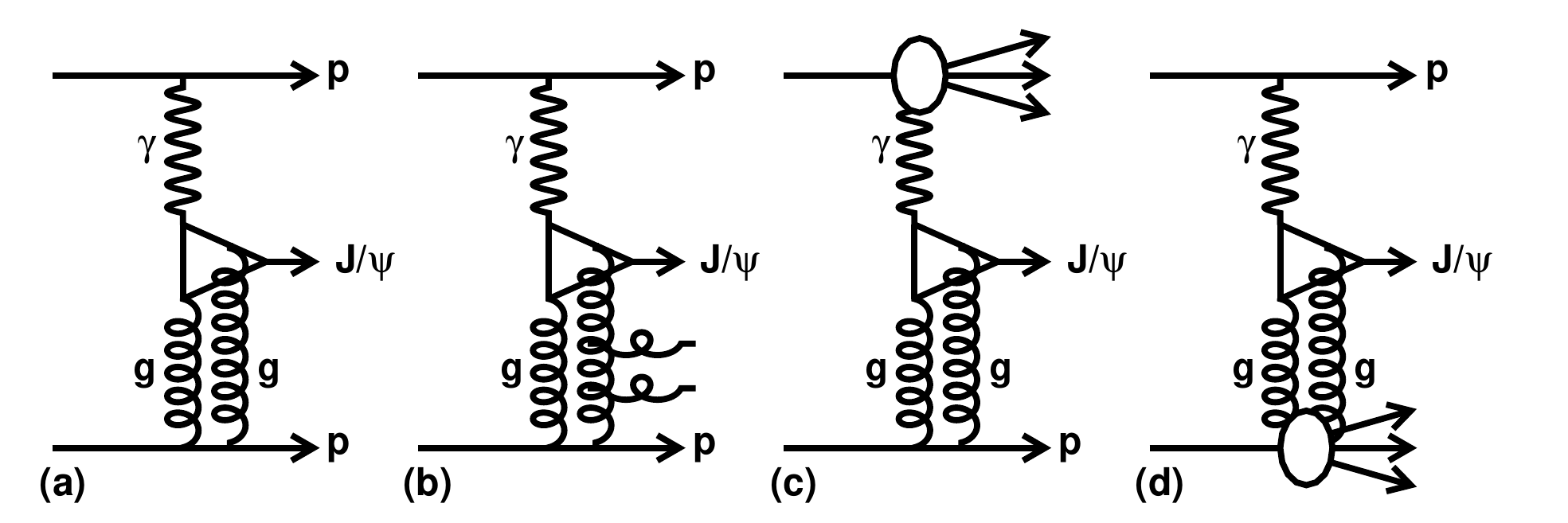}
     \vspace*{-1.0cm}
\end{center}
\caption{
Feynman diagrams displaying (a) exclusive $\jpsi$ photoproduction and (b) inelastic $\jpsi$ photoproduction where a small number of additional particles
are produced due to  gluon radiation and (c,d) proton dissociation.
}
\label{fig:prod}
\end{figure}

\subsection{Inelastic background determination}

The requirement of two tracks and no other visible activity enriches the sample in exclusive events.  
However, this does not guarantee that there is no other activity in the regions outside the $\lhcb$ acceptance.  
The contributions from two non-exclusive processes have been considered: inclusive prompt charmonium produced through colour strings, which leads to
large numbers of additional particles; 
and inelastic $\jpsi$ photoproduction (as shown in Fig.~\ref{fig:prod}), where gluon radiation or proton
dissociation lead to a small number of additional particles.
The former has been evaluated using simulated samples of prompt charmonium,
generated using PYTHIA~\cite{Pythia} and normalised using a data sample where the zero backward tracks requirement has been removed; it is found to be negligible.

\begin{figure}
\begin{center}
  \includegraphics[width=0.7\linewidth]{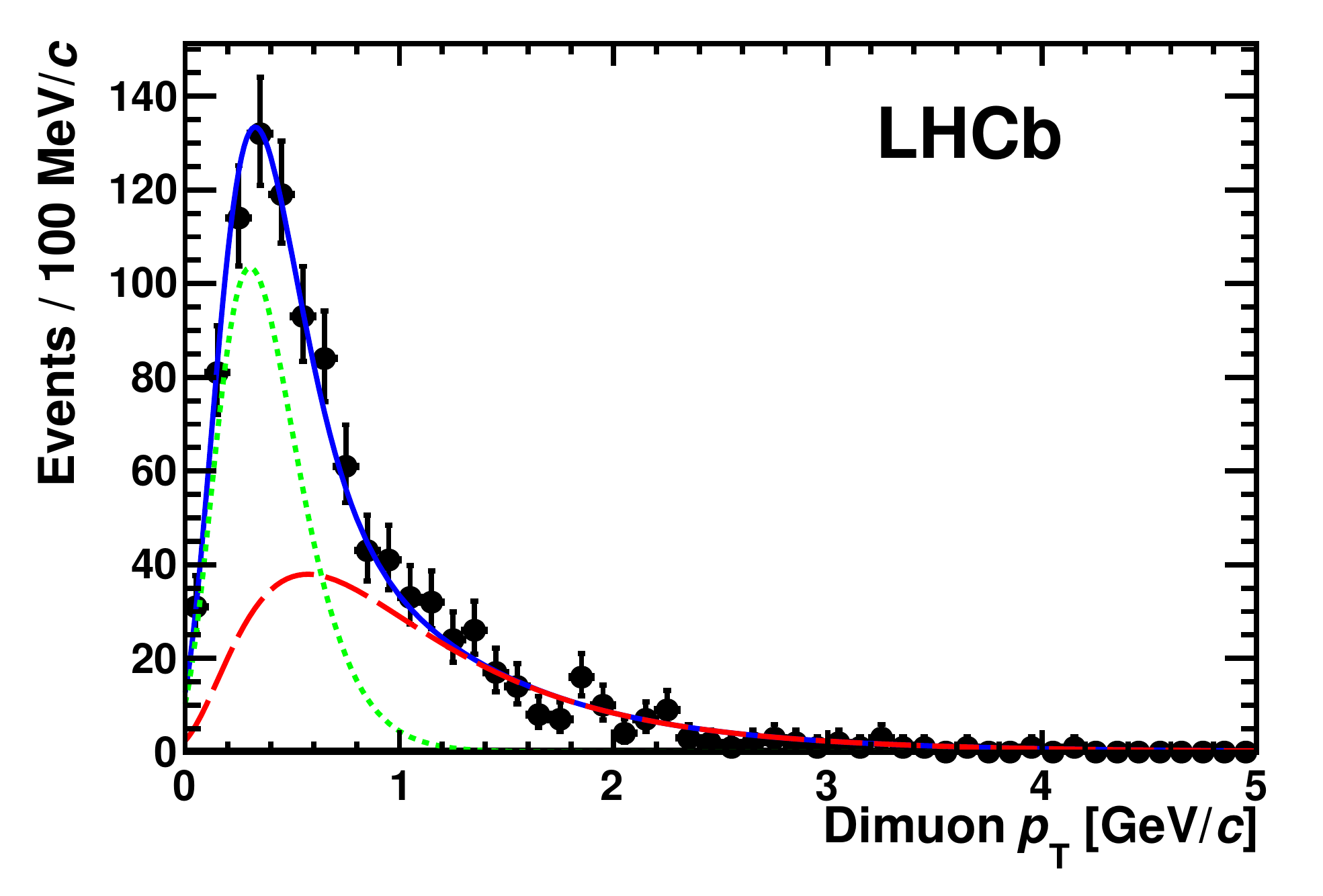}
     \vspace*{-1.0cm}
\end{center}
\small
\caption{
Transverse momentum distribution for exclusive $\jpsi$ candidates with exactly two tracks and no photons. 
The points represent the data. The fit contains an exclusive signal component (short-dashed green curve) as estimated 
by \textsc{SuperChic} and an inelastic background component (long-dashed red curve) as estimated from data. 
}
\label{fig:JPsiPurity}
\end{figure}

The inelastic photoproduction of $\jpsi$ mesons is the dominant background in this analysis. Most of the additional particles produced in these events escape down the beam pipe and thus the inelastic production cross-section as a function of track multiplicity is expected to peak at two tracks. An extrapolation of the inelastic production cross-section from higher track multiplicities is not possible as no reliable simulation is available. Instead, this background is determined from a fit to the $\pt$ spectrum of the exclusive candidates. 
The signal shape is taken from simulation while the background distribution is estimated from data as described below. 
The result is shown in Fig.~\ref{fig:JPsiPurity} for events selected by the software trigger, which has no restriction on the dimuon $\pt$.
The agreement between the predictions and the data is good. 
To reduce the inelastic background contribution the dimuon $\pt$ is required to be less than 900$\mevc$. 
Below this value the fit estimates that (70 $\pm$ 4)$\%$ of the events are exclusive.

The signal shape is taken from the
dimuon $\pt$ distribution of the \textsc{SuperChic} simulation which
assumes a $\pt^{2}$ distribution of the form $\exp(-b \pt^{2})$. 
Regge theory predicts~\cite{Regge} that the slope, $b$, of the $\pt^2$ distribution increases with $W$ according to 
\begin{equation}
b=b_0+4\alpha^\prime\ln{W\over W_0},
\label{eq:bvalue}
\end{equation}
where fits to H1 and ZEUS data~\cite{cH1b, zeus}
give values 
of $b_0$ = 5 GeV$^{-2}c^{2}$, $\alpha^\prime=0.2$ GeV$^{-2}c^{2}$ at a scale $W_0=90$ GeV. 
Producing an object of mass $m_{V}$ at a rapidity $y$ accesses $W$ values of
\begin{equation}
(W_\pm)^2=m_{V}\sqrt{s}\exp(\pm |y|),
\label{eq:wfromy}
\end{equation}
where the two solutions, $W_+$ and $W_-$, correspond to the photon being either an emitter or a target. 
The total cross-section has contributions coming from both $W_+$ and $W_-$, the relative amounts of 
which can be determined from the photon energy spectrum of the proton given in Ref.~\cite{drees}. 
To determine an appropriate $b$ value to describe the $\pt$ spectrum in LHCb, the rapidity range $2.0<y<4.5$ is split into ten equally sized bins and $W_{\pm}$ values are calculated for each bin using Eq.(\ref{eq:wfromy}). Two corresponding $b$ values are found using Eq.(\ref{eq:bvalue}) and then weighted according to the photon energy spectrum.
The mean and root mean square of $b$ in the rapidity bins are 6.1 GeV$^{-2}c^{2}$ and 0.3 GeV$^{-2}c^{2}$, respectively.  
These values are consistent with the results of the fit described below where $b$ is left as a free parameter.

\begin{figure}[h]
\begin{center}$
\begin{array}{cc}
 \subfigure{
            \label{fig:img1}
\includegraphics[width=0.45\linewidth]{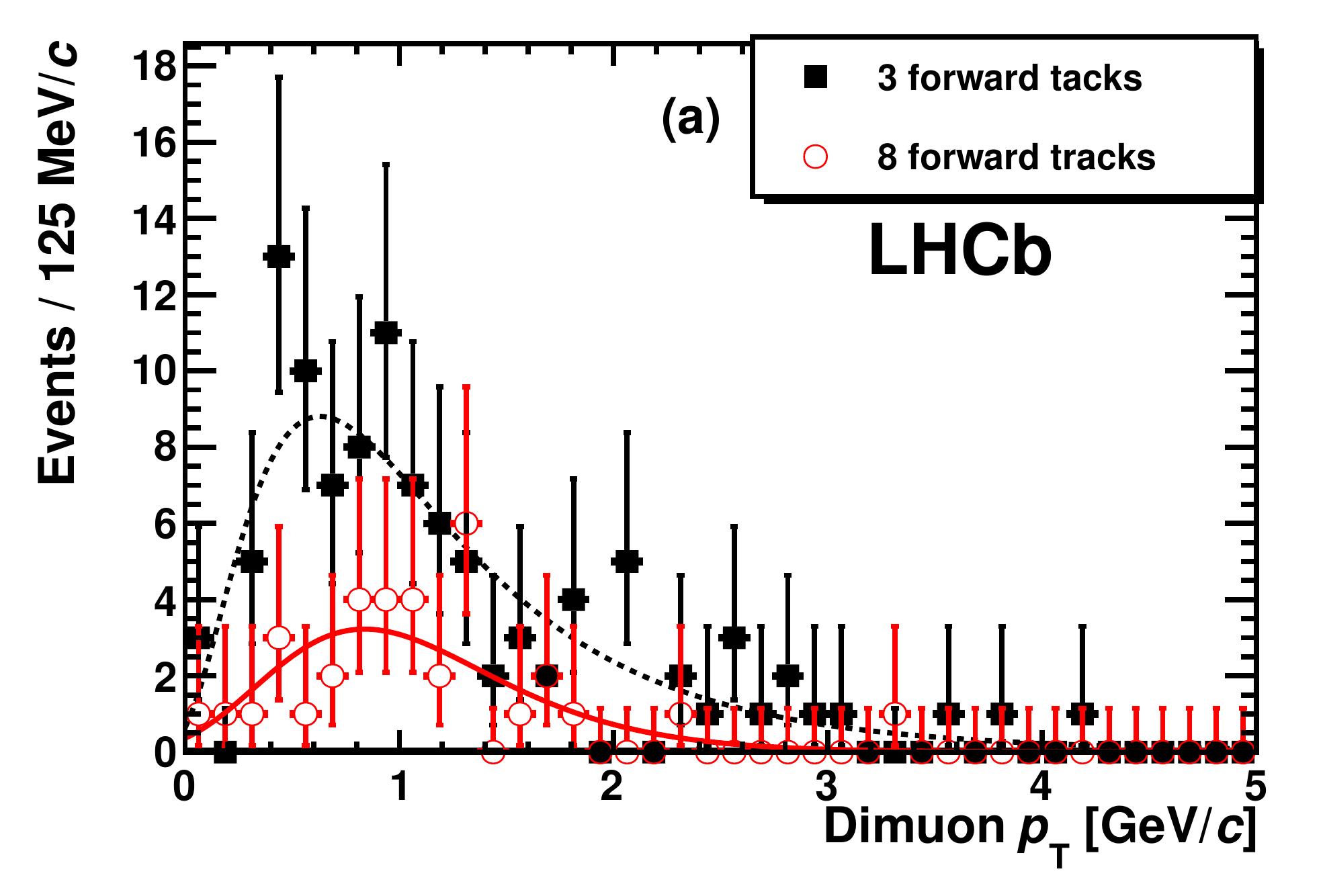} 
} &
 \subfigure{
            \label{fig:img2}
\includegraphics[width=0.45\linewidth]{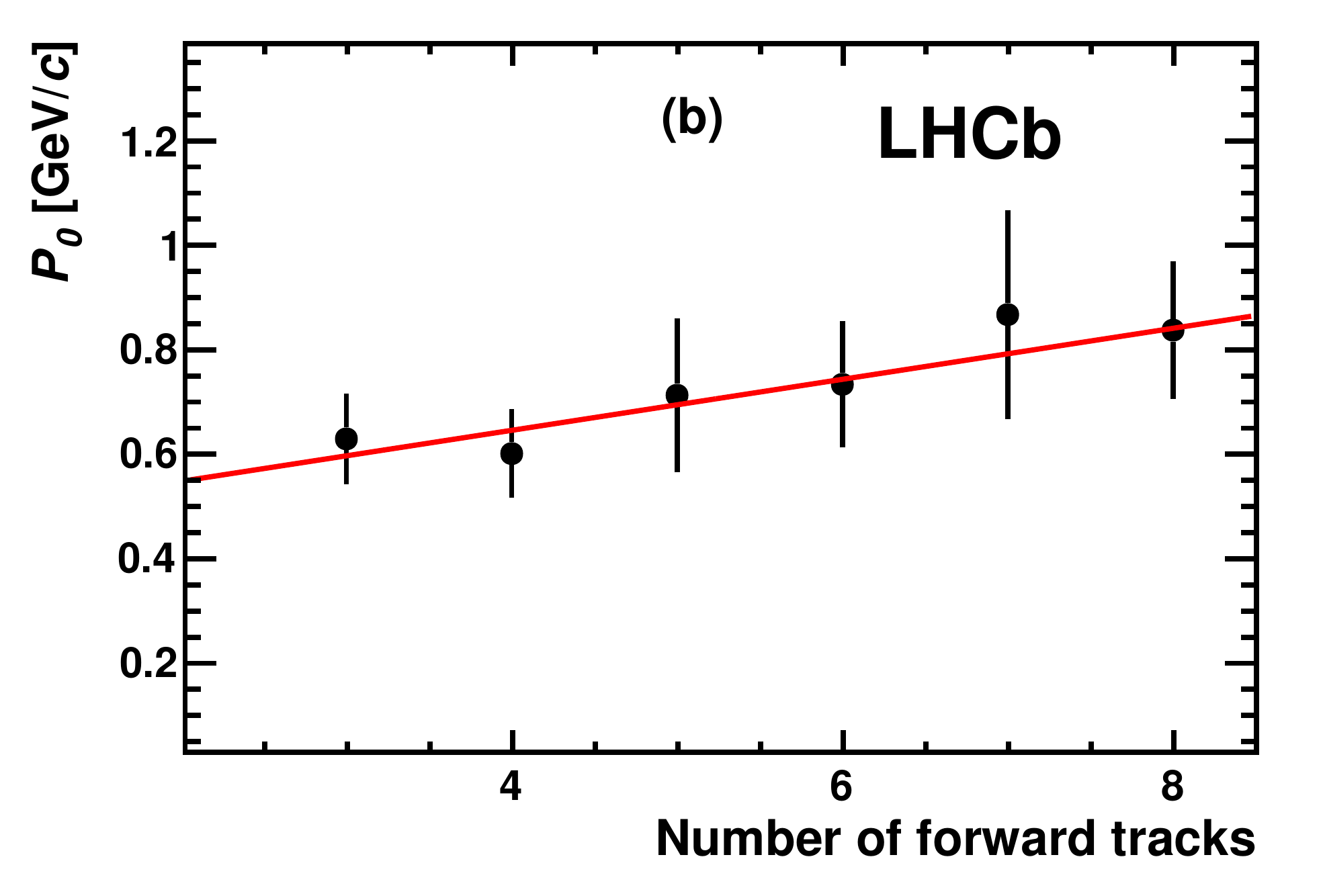}  
}   \\
 \subfigure{
            \label{fig:img1}
\includegraphics[width=0.45\linewidth]{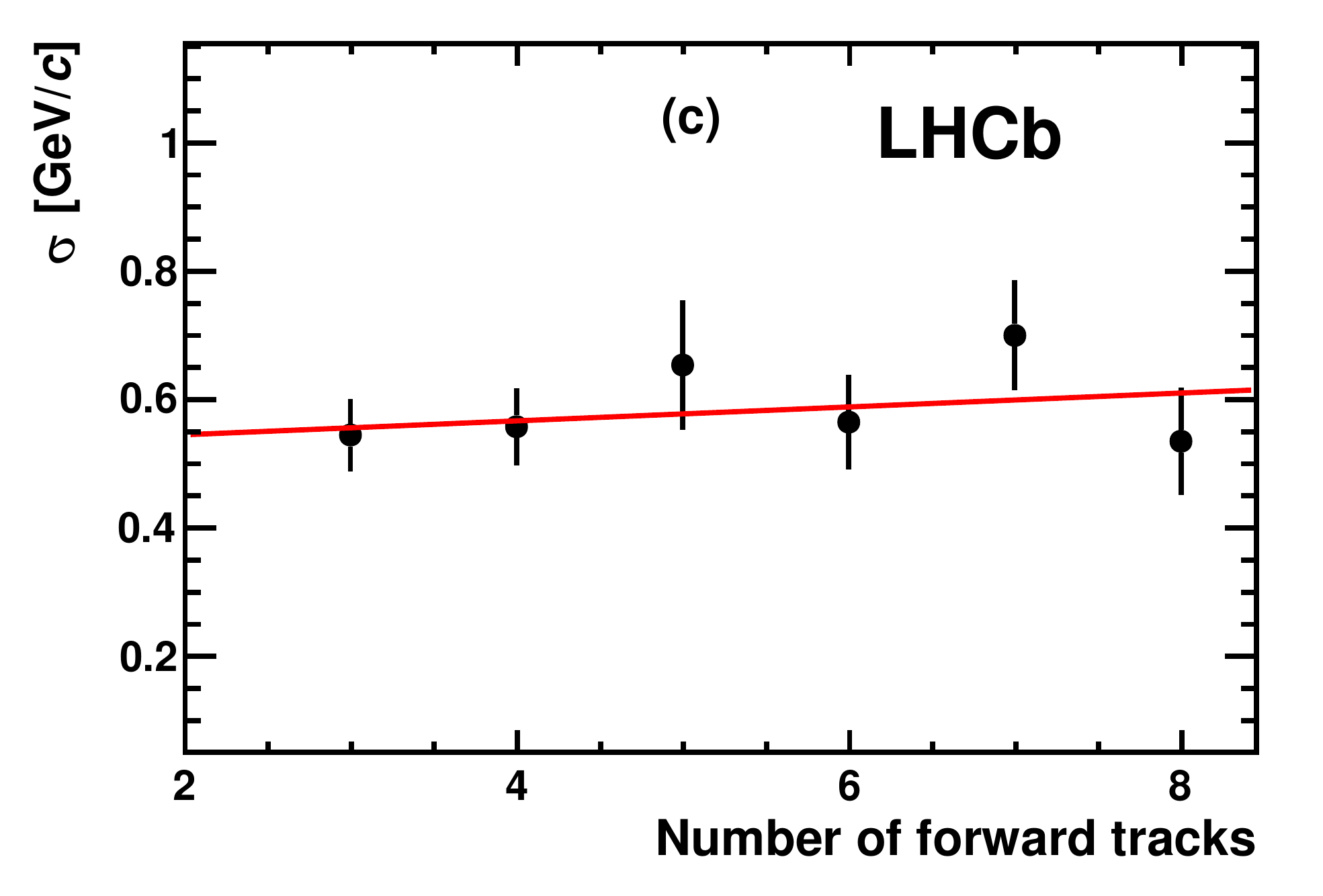} 
}&
 \subfigure{
            \label{fig:img1}
\includegraphics[width=0.45\linewidth]{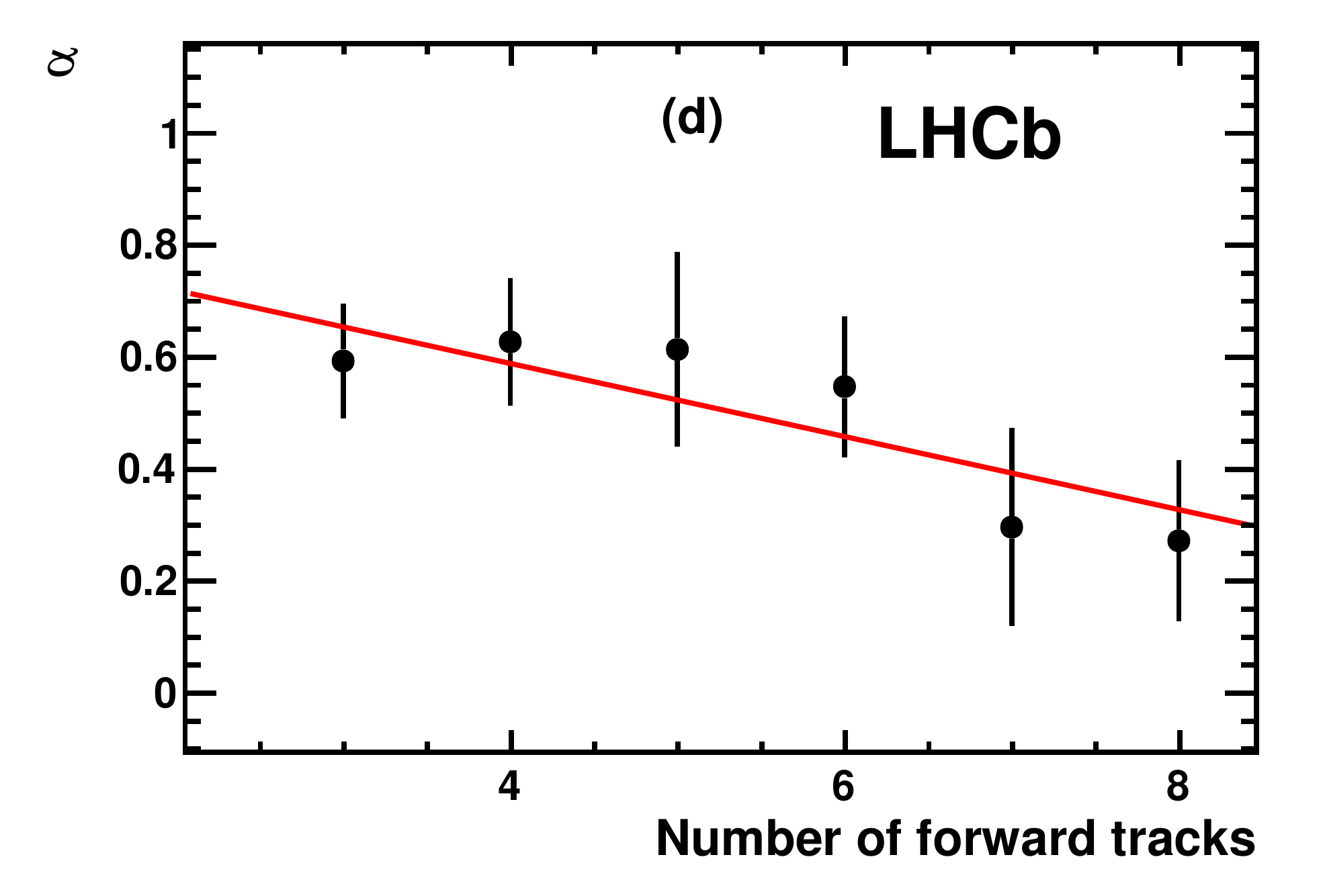} 
} \\
 \vspace*{-1.0cm}
\end{array}$
\end{center}
\caption{
(a) Transverse momentum distribution in $\jpsi$ events with no backward tracks and no photons when there are three (black squares) and eight (red open circles) forward tracks. The results of fitting a Novosibirsk function to the three and eight forward track data are represented by the dashed black and the full red curves respectively.
The values for (b) $P_{0}$, (c) $\sigma$ and (d) $\alpha$ of the $\jpsi$ $\pt$ distribution are shown as a function of the number of forward tracks.
}
\label{fig:JPsiPtVsTrack}
\end{figure}

The shape for the inelastic background is taken from data. 
Non-exclusive candidates are selected by requiring events to contain more than two tracks in the forward region.
The $\pt$ distributions for events with 3, 4, 5, 6, 7 and 8 forward tracks are each fitted separately 
with a Novosibirsk function\footnote{
The Novosibirsk function is defined as
\newline
$N(P;P_{0}, \sigma, \alpha )=A \exp(-0.5(\ln^{2}(1 + \Lambda\alpha(P-P_{0}) )/\alpha^{2} + \alpha^{2}))$
\newline
where $\Lambda = \sinh(\alpha \sqrt{\ln4})/(\sigma\alpha\sqrt{\ln4})$, $P_{0}$ is the peak position, $\sigma$ is the width and $\alpha$ is the tail parameter. },
which is a skewed Gaussian function with three parameters for the peak position ($P_{0}$), width ($\sigma$) and tail ($\alpha$).

Figure~\ref{fig:JPsiPtVsTrack}(a) displays two of the fitted $\pt$ spectra for the $\jpsi$ meson in events with no backward tracks and no photons when there are 3 and 8 forward tracks.
The values for $P_{0}$, $\sigma$ and $\alpha$ are extracted from each fit and plotted as a function of the number of forward tracks in Fig.~\ref{fig:JPsiPtVsTrack}(b), (c) and (d), respectively. 
A linear extrapolation of each parameter is made to predict the inelastic background shape for events with two forward tracks giving 
$P_{0}=0.55 \pm 0.09 \gevc$, $\sigma=0.56 \pm 0.06 \gevc$ and $\alpha=0.72 \pm 0.11$. 

Although a full theoretical prediction is not possible~\cite{Phenom} a dependence of the $\pt$ peak position on the number of tracks 
is expected from general kinematic considerations; additional gluon radiation
will impart a greater $\pt$ to the $\jpsi$ as well as giving extra tracks inside $\lhcb$, while proton
disintegration, leading to additional tracks, is also more likely to occur at higher $\pt$.
This is found to be the case for a number of  related processes using simulated events:  single and central diffractive events in PYTHIA;
the diphoton production of muon pairs with similar masses to the $\jpsi$ in which one or both of the protons disintegrate, as described by LPAIR~\cite{Lpair};
and minimum bias events in PYTHIA. A linear dependence is also observed in $Z\rightarrow\mumu$ events in data although the
average $\pt$ is one order of magnitude larger than in exclusive $\jpsi$ production.

Systematic uncertainties on the inelastic background determination due to both the signal and background shapes have been assessed. 
To determine the former, the fit to the data is repeated leaving $b$ as a free parameter, which returns a value of 
\mbox{5.8 $\pm$ 0.8 GeV$^{-2}c^{2}$}. 
A variation in the estimated percentage of exclusive events of 4$\%$ is observed when $b$ is changed by $\pm 1$ GeV$^{-2}c^{2}$. 
The uncertainty due to the background shape is found by varying the parameters of the phenomenological shape within their statistical uncertainties resulting in a change of 4$\%$ in the estimated fraction of exclusive events.  
Therefore, it is estimated that ($70 \pm 4 \pm 6)\%$ of the $\jpsi$ sample consists of exclusive events, 
where the first uncertainty is statistical and the second is systematic 
due to out understanding of the signal and background shapes.
It is assumed that the $\psitwos$ sample has the same proportion of exclusive events. 

\subsection{Exclusive feed-down background determination} 

Exclusive $\chic$ production can feed down to give a fake exclusive $\jpsi$,
via the $\chic \rightarrow \jpsi \gamma$ decay, when the photon is very soft or is outside the detector acceptance.
There is no corresponding resonance above the $\psitwos$. 
Exclusive $\chic$ candidate events are identified in the data as those containing a $\jpsi$ and a single photon. The background from $\chic$ feed-down is then estimated by scaling the number of observed $\chic$ candidates by the ratio of fake exclusive $\jpsi$ to exclusive $\chic$ candidates in simulated $\chic$ events. 
The feed-down from $\chic$ decay is estimated to account for (9.0 $\pm$ 0.8)$\%$ of the exclusive $\jpsi$ candidates, where the uncertainty includes a contribution due to the uncertainty on the photon reconstruction efficiency in simulation.  

Exclusive $\psitwos$, in particular $\psitwos \rightarrow \jpsi + X$, can also feed down to give a fake exclusive $\jpsi$ if the additional particles are undetected. This is estimated from a simulated sample of $\psitwos$ events which has been normalised using the number of observed $\psitwos \rightarrow \mu^{+} \mu^{-}$ events in data. 
The amount of feed-down to the $\jpsi$ from the $\psitwos$ is estimated to be (1.8 $\pm$ 0.3)$\%$. 

\subsection{Selection summary}

\begin{table}[b]
\centering
\caption{Summary of selection requirements.}
\begin{tabular}{lclcl}
Quantity & Requirement \\
\hline
 Dimuon mass    & within 65$\mevcc$  of known value \\
 Dimuon $\pt$ & $\pt <$ 900$\mevc$       \\     
 Muon $\eta$ &  2.0 $ < \eta_{\mu^{\pm}} < $ 4.5  \\   
Number of backward tracks  &  0  \\
Number of forward tracks  &  2 \\
Number of photons &  0   \\                      
\end{tabular}
\label{tab:Sel-sum}
\end{table}

The requirements for the selection of exclusive $\jpsi$ and $\psitwos$ events are summarised in Table~\ref{tab:Sel-sum}.
In total, 1492 exclusive $\jpsi$ and 40 exclusive $\psitwos$ candidate events pass the selection requirements. The overall purities  (including inelastic, non-resonant, and feed-down backgrounds
where appropriate) are  estimated to be $(62 \pm 4\pm 5)$\% for the $\jpsi$ sample and
$(59 \pm 4 \pm 5)$\% for the $\psitwos$ sample.

A cross-section times branching fraction, $\sigma_{V \rightarrow \mu^{+} \mu^{-}} $, is calculated for the $\jpsi$ and the $\psitwos$ using the number of selected events, $N$, and the equation $\sigma_{V \rightarrow \mu^{+} \mu^{-}}  = pN / (\epsilon L)$ where $\epsilon$ represents the efficiency for selecting the events, 
$p$ is the purity of the sample, and $L$ is the luminosity which has been determined
with an uncertainty of 3.5\%~\cite{lumi}.

\section{Efficiency determination} 

The efficiency $\epsilon$ is the product of five components, 
$\epsilon_{\mathrm{trigger}} \times \space \epsilon^2_{\mathrm{track}} \times \epsilon^2_{\mathrm{muon}} \times \epsilon_{\mathrm{sel}} \times \epsilon_{\mathrm{single}}$ where:
$\epsilon_{\mathrm{trigger}}$ is the efficiency for triggering on events that pass the offline selection; $\epsilon_{\mathrm{track}}$ is the efficiency for reconstructing a track within the fiducial region of the measurement; $\epsilon_{\mathrm{muon}}$ is the efficiency 
for identifying a track as a muon; $\epsilon_{\mathrm{sel}}$ is the efficiency of the selection requirements in the kinematic range of the measurement; and 
$\epsilon_{\mathrm{single}}$ is the efficiency for selecting single interaction events. 
The first four  components have been determined from simulation.
The fifth component accounts for the fact that the selection requirements reject signal events
that are accompanied by a visible proton-proton interaction in the same beam crossing.

The number 
of visible proton-proton interactions per beam crossing, $n$, 
is assumed to follow a Poisson distribution, $P(n)=\mu^{n}\exp(-\mu)/n!$, where $\mu$ is the average number of visible interactions. 
The probability that a signal event is not rejected due to the presence of another visible interaction is given by $P(0)$ and, therefore, $\epsilon_{\mathrm{single}}=\exp(-\mu)$.
This has been calculated throughout the data-taking period in roughly hour-long intervals.
Variations in $\mu$ during this interval have been studied and found to have a negligible effect.
The spread in the value of $\mu$ for different crossing bunch-pairs is small
and its effect is neglected.
The impact of detector noise has also been investigated using data taken when either one beam or no beam circulated and a systematic uncertainty of 0.7$\%$ is evaluated leading to a determination
for  $\epsilon_{\mathrm{single}}$ of $(21.1 \pm 0.1)\%$. 

A systematic uncertainty of  4$\%$ is assigned to $\epsilon_{\mathrm{trigger}}$.
This is based on the difference between the value measured in data and in simulation as determined in Ref.~\cite{cINCJPSI}.

A tag-and-probe technique with $\jpsi \rightarrow \mu^{+} \mu^{-}$ decays~\cite{cTrEff1}
has been used to study $\epsilon_{\mathrm{track}}$.   
The estimated value from this method is found to agree within 1$\%$ with the simulation. 
This value is taken as a systematic uncertainty per track. 

A systematic uncertainty of 2.5$\%$ per muon
is assigned to the determination of $\epsilon_{\mathrm{muon}}$. 
This is based on the results from Ref.~\cite{cINCJPSI} using a tag-and-probe technique. 

The fraction of exclusive $\jpsi$ mesons below the $\pt$ selection threshold of 900$\mevc$ is found to vary by 1$\%$ when $b$, the $\pt$ shape parameter, is changed by $\pm$ 1 GeV$^{-2}c^{2}$. Thus a systematic uncertainty of 1$\%$, due to the uncertainty in the $\pt$ shape,
is assigned to the determination of $\epsilon_{\mathrm{sel}}$.
No systematic uncertainty is assigned due to the  $\jpsi$ polarisation; it is assumed to be transversely polarised due to s-channel helicity conservation.
A summary of the systematic uncertainties of the analysis is shown in Table~\ref{tab:Sys-sum}. 

\begin{table}
\centering
\caption{Relative systematic uncertainties on the measurement.}
\begin{tabular}{lp{6.0cm}lp{6.0cm}l}
  Source & Uncertainty ($\%$) \\
\hline
Luminosity  & 3.5   \\
Trigger efficiency & 4   \\
Tracking efficiency & 2  \\
Identification efficiency   & 5 \\
Selection efficiency    & 1   \\      
Single interaction efficiency & 0.7   \\           
 $\psitwos$ background ($\jpsi$ analysis)   & 0.3 \\     
  $\chic$ background ($\jpsi$ analysis)   & 0.8  \\                  
Signal shape of dimuon $\pt$ fit & 6 \\   
Background shape of dimuon $\pt$fit & 6 \\                                                           
\end{tabular}
\label{tab:Sys-sum}
\end{table}

\section{Results}

The cross-section times branching fraction to two muons with pseudorapidities
between 2.0 and 4.5 is measured for exclusive $\jpsi$ and $\psitwos$
to be
\begin{equation*}
\sigma_{pp\rightarrow \jpsi (\rightarrow \mu^{+} \mu^{-})} (2.0 <\eta_{\mu^{\pm}}< 4.5) = 307 \pm 21 \pm 36~\text{pb},   
\end{equation*}
\begin{equation*}
\sigma_{pp\rightarrow \psitwos (\rightarrow \mu^{+} \mu^{-})} (2.0 <\eta_{\mu^{\pm}}< 4.5) = 7.8 \pm 1.3 \pm 1.0~\text{pb},   
\end{equation*}
\noindent where the first uncertainty is statistical and the second is systematic.

These results are compared to a number of predictions for exclusive production in Table~\ref{tab:CS-sum}.  The predictions for \textsc{Starlight} and \textsc{SuperChic} have been determined using samples of generated events with a full LHCb simulation. 
The other predictions are obtained by 
scaling the differential cross-section in rapidity for each model by an acceptance factor
corresponding to the fraction of mesons at a given rapidity that have both muons
in the fiducial volume, as determined using \textsc{Starlight}.  
For the models of Motyka and Watt, and Gon\c{c}alves and Machado, a rescattering correction
of 0.8 has been assumed~\cite{schaefer}.
The prediction of Sch\"afer and Szczurek is significantly higher than the data; good agreement is
observed with all other predictions.

\begin{table}[b]
\centering
\caption{Comparison of cross-section times branching fraction measurements (pb) with theoretical predictions.}
\begin{tabular}{lp{4.0cm}lp{4.0cm}l}
 Predictions  &  $\sigma_{pp\rightarrow \jpsi (\rightarrow \mu^{+} \mu^{-})}$ &  $\sigma_{pp\rightarrow \psitwos (\rightarrow \mu^{+} \mu^{-})}$  \\
 \hline
Gon\c{c}alves and Machado & 275 \\
\textsc{Starlight} &  292  & 6.1  \\   
Motyka and Watt  & 334  & \\
 \textsc{SuperChic}  & 396 & \\                            
Sch\"afer and Szczurek  & 710 & 17 \\        
\hline        
LHCb measured value & $307 \pm 21 \pm 36$  & $7.8 \pm1.3 \pm 1.0$ \\        
\end{tabular}
\label{tab:CS-sum}
\end{table}

\begin{table}[t]
\begin{center}
\caption{Cross-section measurements (nb) as a function of $\jpsi$ rapidity.}
\begin{tabular}{lp{3.cm}lp{3.cm}lp{3.cm}lp{3.cm}lp{3.cm}l}
Rapidity & 2.00-2.25 & 2.25-2.50 & 2.50-2.75 & 2.75-3.00 \\
\hline 
${\mathrm{d}\sigma\over\mathrm{d}y} (\jpsi)$ & 3.2 $\pm$ 0.8 $\pm$ 0.9 & 4.5 $\pm$ 0.5 $\pm$ 0.8 & 5.3 $\pm$ 0.4 $\pm$ 0.9 & 4.4 $\pm$ 0.3 $\pm$ 0.7 \\

\\
Rapidity & 3.00-3.25 & 3.25-3.50 & 3.50-3.75 & 3.75-4.00 \\
\hline 
${\mathrm{d}\sigma\over\mathrm{d}y} (\jpsi)$ & 5.5 $\pm$ 0.3 $\pm$ 0.8  
& 4.8 $\pm$ 0.3 $\pm$ 0.7 & 5.2 $\pm$ 0.3 $\pm$  0.8 & 4.8 $\pm$ 0.4 $\pm$ 0.8 \\
\\
Rapidity &  4.00-4.25 & 4.25-4.50 \\
\hline 
${\mathrm{d}\sigma\over\mathrm{d}y} (\jpsi)$ 
& 4.7 $\pm$ 0.5 $\pm$ 0.9 
& 4.1 $\pm$ 0.9 $\pm$ 1.3 \\

\end{tabular}
\end{center}
\label{tab:CSvsRapiditySum}
\end{table}

Combining the vector meson branching fractions to two muons with acceptance factors determined from \textsc{Starlight} gives a ratio of $\psitwos$ to $\jpsi$ production of 0.19 $\pm$ 0.04.
This can be compared to a value of 0.16 according to \textsc{Starlight} and about 0.2 
according to Sch\"afer and Szczurek. CDF measured this ratio to be 0.14 $\pm$ 0.05~\cite{cCDF} and at HERA it was measured to be 0.166 $\pm$ 0.012~\cite{cH1b, zeus} although these were at different values of $W$.

The differential $\jpsi$ cross-section is also measured in ten bins of $\jpsi$ rapidity.
The trigger and selection efficiencies are calculated from the simulation in bins of rapidity.
The systematic uncertainties are dominated by the purity and the statistical uncertainty coming from the size of the simulation sample.
The purity within each bin is assumed to be the same as in the integrated sample.  
An acceptance factor in each rapidity bin has been calculated using the \textsc{Starlight} simulation. 
Table~\ref{tab:CSvsRapiditySum} summarises the differential cross-section result. 

The present results can be compared to H1 and ZEUS results~\cite{cH1b, zeus} for the photoproduction of 
$\jpsi$. This is possible as the underlying production mechanism is the same: at HERA the photon radiates from an electron,
while at the LHC the photon radiates from a proton. 
The differential cross-section for proton-proton exclusive photoproduction of a vector meson with mass $m_{V}$ can be obtained by weighting
the photon-proton exclusive production cross-section by 
the photon flux, ${d}n/{d}k$, for a photon of energy $k$~\cite{klein04,cMW}

\begin{equation}
{{d}\sigma\over {d}y}_{pp\rightarrow pVp}=
r(y)\biggl[
k_+{dn\over dk_+}\sigma_{\gamma p\rightarrow Vp}(W_+)
+k_-{dn\over dk_-}\sigma_{\gamma p\rightarrow Vp}(W_-)
\biggr],
\label{eq:cs}
\end{equation}
\begin{equation}
k_\pm\approx (m_{V}/2) \exp(\pm |y|), 
\label{eq:k}
\end{equation}

\noindent where $W_{\pm}$ is defined as in Eq.(\ref{eq:wfromy}) and $r(y)$ is an absorptive correction which depends on $y$~\cite{schaefer}. 

Assuming the validity of a power law dependence of the form $aW^\delta$ to describe 
$\sigma_{\gamma p\rightarrow Vp}$, the proton-proton differential cross-section can
be written as
\begin{equation}
{d\sigma\over dy}_{pp\rightarrow pVp}
=
a(2\sqrt{s})^{\delta/2}
r(y)\biggl[ 
 {dn\over dk_+} {k_+}^{1+\delta/2}
+ {dn\over dk_-} {k_-}^{1+\delta/2}
\biggr].
\label{eq:power}
\end{equation}

The parameters for the power law dependence of the photoproduction cross-section are
found by fitting the differential cross-section data from Table~\ref{tab:CSvsRapiditySum} with the functional form given in Eq.(\ref{eq:power}).
The uncertainties between bins are taken to be uncorrelated for all sources except for the
purity, which is fully correlated between bins.  
For the description of ${dn\over dk}$, the photon energy spectrum given in Ref.~\cite{drees} is used.
The absorptive corrections have been calculated in Ref.~\cite{schaefer} for proton-proton collisions and have
a value of 0.85 at $y=0$ and 0.75 at $y=3$ with a rather flat dependence on $y$.
This analysis assumes a shape $r(y)=0.85-0.1|y|/3$.
The fit to the data in Table~\ref{tab:CSvsRapiditySum} gives 
values of $a = 0.8^{+1.2}_{-0.5}$ nb and $\delta=0.92\pm 0.15$ with a $\chi^2$
of 4.3 for 8 degrees of freedom, indicating the results are consistent with the hypothesis
of a power law dependence. The values obtained 
are also consistent with the results from HERA, albeit with much larger uncertainties.

\subsection{Evaluation of the photon-proton cross-section}

\begin{figure}[t]
\begin{center}$
\begin{array}{c}
 \includegraphics[width=0.85\linewidth]{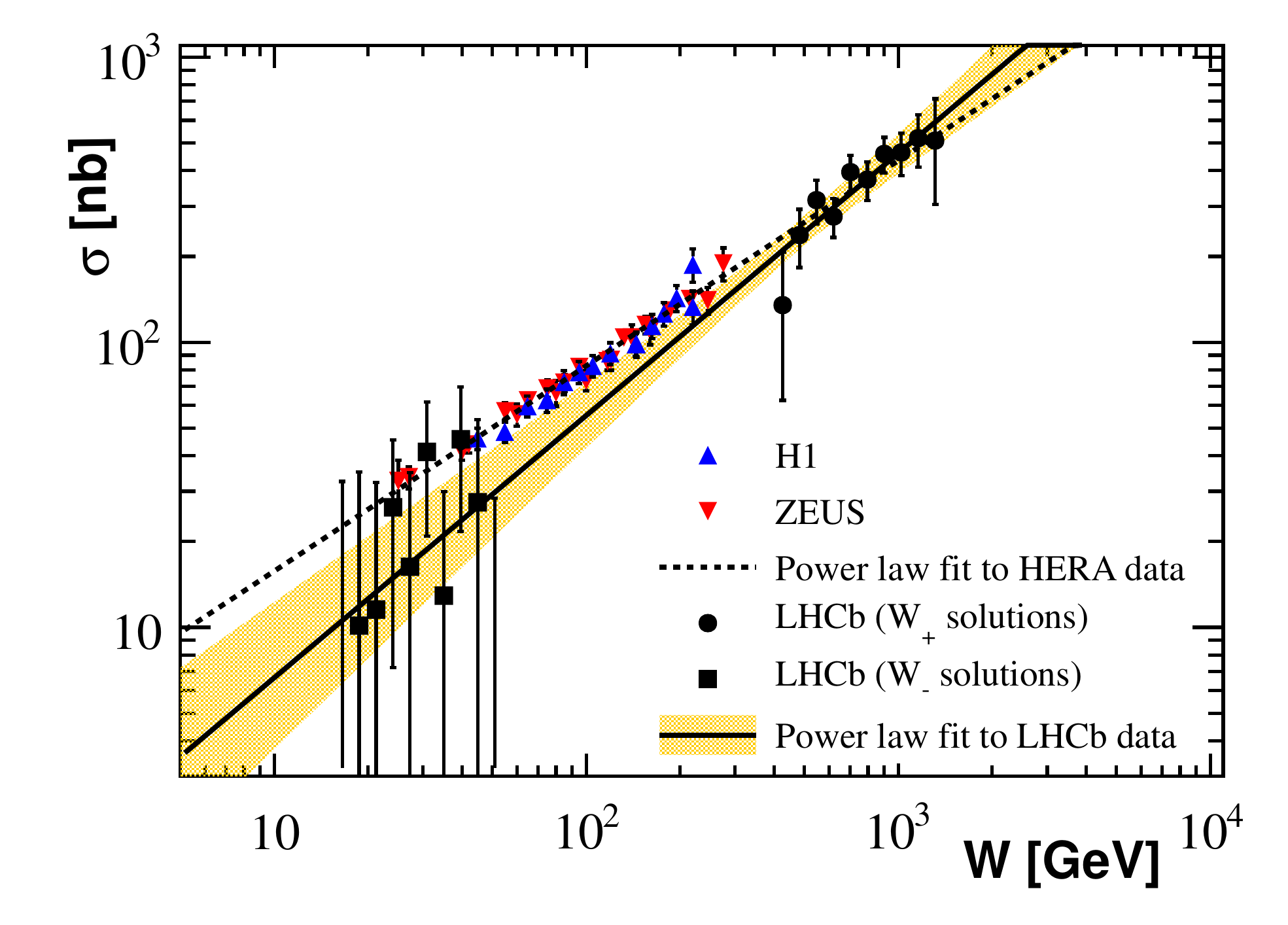}
 \vspace*{-1.0cm}
\end{array}$
\end{center}
\small
\caption{
Dependence of $\jpsi$ photoproduction cross-section on the centre-of-mass energy of the photon-proton system.
The blue (red) triangles represent the data from H1 (ZEUS)~\cite{cH1b,zeus}.  
The black dots and squares are derived from the LHCb differential cross-section as a function of rapidity.
The dashed and full lines are the power law dependences determined from the HERA and LHCb data, respectively.
The uncertainty on the LHCb power law determination is shown by the shaded band.
}
\label{fig:lhcbhera}
\end{figure}

The differential cross-sections for the process $pp\rightarrow p\jpsi p$
given in Table~\ref{tab:CSvsRapiditySum} are transformed into cross-sections
for the process $\gamma p\rightarrow \jpsi p$ using a re-arrangement of Eq.(\ref{eq:cs})
 \begin{equation}
\sigma_{\gamma p\rightarrow Vp}(W_{\pm})={
{1/ r(y)}{d\sigma \over dy}_{pp\rightarrow pVp} -
k_{\mp}{dn \over dk_{\mp}}\sigma_{\gamma p\rightarrow Vp}(W_{\mp})
\over
k_{\pm}{dn \over dk_{\pm}}
}.
\label{eq:cs1}
\end{equation}

The photoproduction cross-sections at $W_{+}$ and $W_{-}$ are determined independently using Eq.(\ref{eq:cs1}) and substituting into the right-hand side the expected cross-section for the alternative $W$ solution from the power law determined above.  

The LHCb data are plotted together with the H1 and ZEUS data in Fig.~\ref{fig:lhcbhera}. 
The power law dependences determined from LHCb data and from HERA data are also indicated.
The uncertainty on the LHCb power law determination is shown by the shaded band.
Only experimental uncertainties are shown on the LHCb data, which have significant bin-to-bin correlations due to the purity determination.  
Theoretical uncertainties that account for the power law assumption and the absorptive correction are not included. 
Both are smaller than the current experimental uncertainties;
using the HERA power law in place of the LHCb power law in Eq.(\ref{eq:cs1})  
changes the estimated values of $\sigma_{\gamma p\rightarrow Vp} $ by about half the experimental uncertainty, while changing the absorptive correction by 5$\%$ changes 
 $\sigma_{\gamma p\rightarrow Vp} $ by about one quarter of the experimental uncertainty. 
With the precision of the current data, the LHCb results are consistent with HERA and confirm a similar power law behaviour for the photoproduction cross-section.

\section{Conclusion}

The first observations of exclusive $\jpsi$ and $\psitwos$ production in proton-proton collisions
have been made.
The cross-sections times branching fraction to two muons with pseudorapidities
between 2.0 and 4.5 are measured to be
$
307 \pm 21 \pm 36~\text{pb}
$
and
$
7.8 \pm 1.3 \pm 1.0~\text{pb}
$
for exclusive $\jpsi$ and $\psitwos$, respectively

The measured cross-sections are in agreement with the theoretical predictions of \textsc{Starlight}, \textsc{SuperChic}, Gon\c{c}alves and Machado, and Motyka and Watt. 
The differential cross-section for $\jpsi$ production as a function of rapidity
has also been measured. This has allowed the $\jpsi$ photoproduction cross-section as a function of the photon-proton centre-of-mass energy to be determined.
The data are consistent with a power law dependence and the parametric form is in
broad agreement with previous results from H1 and ZEUS.

\section*{Acknowledgements}

\noindent 
We thank Lucian Harland-Lang, Valery Khoze, James Stirling and Graeme Watt for many helpful discussions.   
We express our gratitude to our colleagues in the CERN
accelerator departments for the excellent performance of the LHC. We
thank the technical and administrative staff at the LHCb
institutes. We acknowledge support from CERN and from the national
agencies: CAPES, CNPq, FAPERJ and FINEP (Brazil); NSFC (China);
CNRS/IN2P3 and Region Auvergne (France); BMBF, DFG, HGF and MPG
(Germany); SFI (Ireland); INFN (Italy); FOM and NWO (The Netherlands);
SCSR (Poland); ANCS/IFA (Romania); MinES, Rosatom, RFBR and NRC
``Kurchatov Institute'' (Russia); MinECo, XuntaGal and GENCAT (Spain);
SNSF and SER (Switzerland); NAS Ukraine (Ukraine); STFC (United
Kingdom); NSF (USA). We also acknowledge the support received from the
ERC under FP7. The Tier1 computing centres are supported by IN2P3
(France), KIT and BMBF (Germany), INFN (Italy), NWO and SURF (The
Netherlands), PIC (Spain), GridPP (United Kingdom). We are thankful
for the computing resources put at our disposal by Yandex LLC
(Russia), as well as to the communities behind the multiple open
source software packages that we depend on.

\addcontentsline{toc}{section}{References}
\bibliographystyle{LHCb}
\bibliography{main}

\end{document}